# Phase stabilization of Kerr frequency comb internally without nonlinear optical interferometry

S.-W. Huang[1*+], A. Kumar[1+], J. Yang[1], M. Yu[2], D.-L. Kwong[2], and C. W. Wong[1*]

[1] Fang Lu Mesoscopic Optics and Quantum Electronics Laboratory, University of California Los Angeles, CA, USA

[2] Institute of Microelectronics, Singapore, Singapore

[*] swhuang@seas.ucla.edu, cheewei.wong@ucla.edu

[+] these authors contributed equally to this work

Optical frequency comb (OFC) technology has been the cornerstone for scientific breakthroughs such as precision frequency metrology, redefinition of time, extreme light-matter interaction, and attosecond sciences. While the current mode-locked laser-based OFC has had great success in extending the scientific frontier, its use in real-world applications beyond the laboratory setting remains an unsolved challenge. Microresonator-based OFCs, or Kerr frequency comb, have recently emerged as a candidate solution to the challenge because of their preferable size, weight, and power consumption (SWaP). On the other hand, the current phase stabilization technology requires either external optical references or power-demanding nonlinear processes, overturning the SWaP benefit of Kerr frequency combs. Introducing a new concept in phase control, here we report an internally phase stabilized Kerr frequency comb without the need of any optical references or nonlinear processes. We describe the comb generation analytically with the theory of cavity induced modulation instability, and demonstrate for the first time that the optical frequency can be stabilized by control of two internally accessible parameters: an intrinsic comb offset $\xi$ and the comb spacing $f_{rep}$. Both parameters are phase locked to microwave references, with 55 mrad and 20 mrad residual phase noises, and the resulting comb-to-comb frequency uncertainty is 0.08 Hz or less. Out-of-loop measurements confirm good coherence and stability across the comb, with measured optical frequency fractional instabilities of $5 \times 10^{-11}/\sqrt{\tau}$. The new phase stabilization method preserves the Kerr frequency comb's key advantages and potential for chip-scale electronic and photonic integration.

Phase stabilized OFC, with its thousands of coherent and stable spectral lines simultaneously, bridges the two previously independent research frontiers in ultrastable laser physics and ultrafast optical science *(1-5)*. The phase stabilization requires two dimensional feedback controls on the OFC's two degrees of freedom, the comb spacing and one of the comb line optical frequencies. While the comb spacing can be readily measured with a high-speed photodetector, the assessment of comb line optical frequency fluctuation requires sophisticated experimental implementation. One approach is to compare the OFC against an external optical reference, and previous phase stabilization of Kerr frequency comb has been predominantly demonstrated with schemes based on this approach *(6-9)*. The requirement of an external optical reference, however, limits the achievable compactness of Kerr frequency comb and impairs its potential for chip-scale electronic and photonic integration. The other approach is to devise a nonlinear optical interferometry which reveals the optical frequency instability through the so-called carrier-envelope-offset frequency $f_{ceo}$, an internal OFC property resulting from difference in the phase and group velocities *(1, 2)*.

Figure 1a shows the schematic of a state-of-the-art *f-2f* nonlinear interferometer widely adopted to measure the $f_{ceo}$ *(10)*. First, the output pulse from a mode-locked laser is spectrally broadened in a highly nonlinear photonic crystal fiber such that its optical spectrum spans more than an octave. Then the octave-spanning spectrum is separated into two parts: the lower-frequency end undergoes second-harmonic generation in a nonlinear crystal while the higher-frequency end only experiences free-propagation. Finally, the two beams are put together in both transverse and longitudinal coordinates for them to interfere on a photodetector and generate a beat note at $f_{ceo}$. For the nonlinear processes to work properly, spectral broadening in particular, few-cycle pulses with peak powers in the 10-kW level are required *(1)*.

While the Kerr frequency comb is approaching the performance of mode-locked laser-based OFC in many aspects *(11-23)*, its output pulse duration and peak power are still lower by orders of magnitude. Application of *f-2f* or *2f-3f* nonlinear interferometer techniques to the Kerr frequency comb is thus challenging and power demanding. The pulse duration can potentially be improved by finer dispersion engineering, but the peak power is fundamentally limited by the bandwidth-efficiency product *(24)* and the large comb spacing. On the other hand, the 10 to 100 GHz comb spacing of Kerr frequency comb is considered an advantageous feature for applications like coherent Raman spectroscopy, high bandwidth telecommunication, optical arbitrary waveform generation, and astrospectrograph calibration *(25-27)*. In the recent pioneering demonstration of self-referenced Kerr frequency comb where *f-2f* nonlinear interferometer technique is adopted *(23)*, multiple stages of high power erbium doped fiber amplifiers (EDFAs) are implemented to boost the peak power of the Kerr frequency comb. The approach with cascaded EDFAs is successful, but it also compromises the Kerr frequency comb's SWaP advantage and potential for chip-scale electronic and photonic integration.

Kerr frequency comb is distinct from the mode-locked laser-based OFC in the generation mechanism (Figure 1b). Here, the cw pump laser not only provides the gain for the comb formation but also serves as the $0^{th}$ comb line of the Kerr frequency comb. When the pump power driving the microresonator exceeds a certain threshold, modulation instability (MI) first occurs and four-wave mixing (FWM) then follows, resulting in the formation of primary comb lines separated by $\Delta$ and secondary comb lines spaced by $f_{rep}$ *(11)*. In general, $\Delta$ is incommensurate with $f_{rep}$ and thus such Kerr frequency comb exhibits an intrinsic offset frequency, $\xi$, that can be detected directly in a photodetector. With the proper choice of pump power $P_p$ and pump frequency $f_p$, the Kerr frequency comb with only one set of primary comb lines can be generated. Consequently we can achieve a $\xi$ that is *uniquely* defined (other types of Kerr frequency comb are shown in Supplementary Information section I). Figure 1c shows the optical and electrical spectra of an example Kerr frequency comb in the telecommunication C-band wavelength range. Formation of primary comb families and overlap between secondary comb lines are observed (left inset), resulting in two distinct electrical beat notes at $f_{rep}$ = 17.9 GHz and $\xi$ = 523.35 MHz (right inset). Measuring the beat notes at various different spectral segments with a tunable 0.22-nm bandpass filter, we confirm the existence of only one primary comb family and the uniformity of $f_{rep}$ and $\xi$ across our Kerr frequency comb (Figure 1d).

In this work, we demonstrate that the unique generation mechanism of Kerr frequency comb, analytically described by the theory of cavity induced modulation instability, provides another phase stabilization scheme in a distinctly different way. Neither external optical references nor power-demanding nonlinear processes are necessary in the new method. After the stabilization of $f_{rep}$, we show that $\xi$ is specifically sensitive to the pump frequency fluctuation, which is at the same time the $0^{th}$ Kerr comb line optical frequency, and thus it is a good internal comb property that can be utilized to gauge the optical frequency instability. Phase locking of $f_{rep}$ and $\xi$ to low noise microwave oscillators guarantees the optical frequency stability of Kerr frequency comb.

We achieve 20 mrad and 55 mrad residual phase noises in both feedback loops, with a resulting comb-to-comb frequency uncertainty of 0.08 Hz or less. Furthermore, out-of-loop characterization confirms the improved optical frequency stability of the internally stabilized chip-scale Kerr frequency comb, measuring an Allan deviation (AD) of $5 \times 10^{-11}/\sqrt{\tau}$ when the gate time is below 5 second. No apparent difference in stability is observed between different spectral sections, demonstrating the good coherence across the comb.

Figure 2a depicts the schematic of the proposed phase stabilization setup where the two degrees of freedom of Kerr frequency comb can be both gauged internally without any power-demanding nonlinear process. Detailed descriptions of the chip fabrication and measurement setup are included in the Method and Supplementary Information sections II and III respectively. The $Si_3N_4$ microresonator is fabricated with CMOS-compatible processes and the spiral design ensures that the relatively large resonator fits into a tight field-of-view to avoid additional cavity losses introduced by photomask stitching and discretization errors. The microresonator is critically coupled with a loaded quality factor approaching 600,000 (intrinsic quality factor at 1,200,000). A 600 μm long adiabatic mode converter is implemented to improve the coupling efficiency from the free space to the bus waveguide. The input-output insertion loss for the whole setup does not exceed 5.5 dB. To isolate the microresonator from ambient thermal fluctuations, the chip is first mounted on a temperature-controlled chip-holder and enclosed in an acrylic chamber. To shield it against acoustic noise and vibrations, the whole enclosure is first placed on a sorbothane sheet and on an active-controlled optical table. The comb spacing of 17.9 GHz is directly measurable by sending the output to a high speed photodetector (bottom inset). The comb spacing is then phase locked and stabilized to a microwave oscillator by controlling the pump power through a fiber electro-optic modulator (primary loop) and the gain of the erbium-doped fiber amplifier (slow loop). Quality of the $f_{rep}$ stabilization is detailed in the Supplementary Information section III, with an r.m.s. phase error, integrated from 6 Hz to 600 kHz, at 20 mrad. Of note, the free-running offset frequency $\xi$ is much noisier than the comb spacing $f_{rep}$ due to the additional multiplier in the constitutive equation:

$$\xi = \Delta - \left\lfloor \frac{\Delta}{f_{rep}} \right\rfloor f_{rep} \qquad (1)$$

To this end, $f_{rep}$ stabilization loop is always engaged before measurements on the offset frequency are conducted. Comparison between free-running and post-$f_{rep}$ stabilization $\xi$ is included in the Supplementary Information section III. As the offset frequency is localized to the spectral region where secondary comb lines overlap, a 0.22 nm optical bandpass filter is used to select the overlapped comb lines around 1553.5 nm for detection. The beat note is thus improved to 50 dB above the noise floor with a resolution bandwidth (RBW) of 10 kHz (top inset), sufficient for a reliable feedback stabilization (more than 45 dB with 10 kHz RBW). The offset frequency is divided by 15 before it is phase locked and stabilized to a microwave synthesizer. The pre-scaling reduces the phase fluctuation, while preserving the fractional frequency instability, and thus it makes the $\xi$ phase-locked loop more robust against noise. The high-bandwidth feedback on $\xi$ is achieved by direct current modulation of the external cavity diode laser (ECDL), and the slow feedback is done through piezoelectric transducer control of the ECDL. Out-of-loop stability of the Kerr frequency comb is evaluated by heterodyne beating with a state-of-the-art fiber laser frequency comb (FFC). The microwave oscillators and frequency counters are all referenced to a rubidium-disciplined crystal oscillator with a frequency fractional instability of $5\times10^{-12}$ at 1 second.

The Kerr frequency comb generation mechanism can be described by the nonlinear Schrödinger equation and the cavity boundary condition *(28)*:

$$\frac{\partial E^n(z,t)}{\partial z} = -\frac{\alpha}{2}E^n(z,t) - i\frac{\beta_2}{2}\frac{\partial^2 E^n(z,t)}{\partial t^2} + i\gamma|E^n(z,t)|^2 E^n(z,t), \qquad (2)$$

$$E^{n+1}(0,t) = \sqrt{1-T}E^n(L,t)exp(i\varphi_0) + \sqrt{T}E_i, \qquad (3)$$

where $E^n(z,t)$ is the electric field envelope function at the $n^{th}$ cavity round-trip, $z$ is the propagation, $t$ is the retarded time, $\alpha$ is cavity round-trip loss, $\beta_2$ is the group velocity dispersion (GVD), $\gamma$ is the nonlinear coefficient, $T$ is transmission coefficient of the coupler, and $\varphi_0$ is the phase accumulated in a round-trip. Here the microresonator is assumed to be critically coupled, for simplicity. Under the mean-field approximation and the good cavity limit, the primary comb spacing, which depends on the optimal frequency where modulation instability gain reaches its maximum, can be solved as (Supplementary Information section IV):

$$\Delta = \frac{1}{\sqrt{\pi c|\beta_2|}}\sqrt{\eta\left(n_g f_p - N\frac{n_g^2}{n_0}f_{rep} - \frac{\gamma c P_{int}}{\pi}\right)} \qquad (4)$$

where $\eta = \frac{\beta_2}{|\beta_2|}$ is the sign of the GVD, $n_g$ is the group index, $n_o$ is the refractive index, $N$ is the longitudinal mode number, $c$ is the speed of light in vacuum, and $P_{int}$ is the intra-cavity pump power. Equations (1) and (4) explicitly show the dependence of $\xi$ on $f_p$, $f_{rep}$, and $P_{int}$. In the high-Q $Si_3N_4$ microresonator, $P_{int}$ is resonantly enhanced to be as high as 30 W and it is the dominant heat source to change the cavity temperature and subsequently the comb spacing *(8)*. For instance, a pump power variation of 0.12 % results in a $1.6 \times 10^{-5}$ fractional change in the comb spacing, corresponding to a large cavity temperature change of 1 K. Thus, we expect the $f_{rep}$ stabilization will effectively eliminate the $P_{int}$ fluctuation. Under this approach, the offset frequency is thereby reduced to solely a function of pump frequency once the comb spacing is stabilized. Feedback controls of $f_{rep}$ and $\xi$ can thus fully stabilize the Kerr frequency comb. Figures 2c and 2d plot the offset frequency as a function of pump frequency and applied pump power, experimentally measured after the $f_{rep}$ stabilization. We observe that the offset frequency remains constant at different applied pump power but scales linearly with the pump frequency at a slope of $3.7 \times 10^{-2}$. The measurements validate our approach that $f_{rep}$ stabilization effectively eliminates the intra-cavity pump power fluctuation and reduces the dependence of $\xi$ to just a function of pump frequency. Mode hybridization in the current multi-mode $Si_3N_4$ microresonator leads to abrupt increase of local GVD and results in the pinning of primary comb lines *(29)*. The effect reduces the slope, *i.e.* sensitivity, of offset frequency in gauging the pump frequency fluctuation (Eq. 4). Nevertheless, the sensitivity is already more than two orders of magnitude larger than the optical frequency division ratio, $\partial f_{rep}/\partial f_{opt} \sim 10^{-4}$, and thus the instability of the Kerr frequency comb $\delta f_{opt} = \frac{1}{\partial f_{rep}/\partial f_{opt}} \delta f_{rep} + \frac{1}{\partial \xi/\partial f_p} \delta \xi$ is still only limited by the residual error and the local oscillator of the $f_{rep}$ stabilization loop.

Figures 3a and 3b show the quality of the $\xi$ stabilization. To minimize the crosstalk between the two phase-locked loops, here the proportional-integral corner frequency is set lower than that of the $f_{rep}$ stabilization loop. On the other hand, a second integrator at 500 Hz is included to better suppress low frequency noise. Compared to the unstabilized beat note shown in the inset of Figure 2a, the stabilized $\xi$ shows a clear resolution limited coherent spike (Figure 3a). The noise oscillation at 205 kHz is the remaining crosstalk derived from the $f_{rep}$ stabilization loop. The single-sideband phase noise of the reference microwave oscillator is plotted in Figure 3b along with the residual loop error. While the low frequency noise is well suppressed to below the reference, excessive phase noise above 2 kHz from carrier is observed. The r.m.s. phase error integrated from 6 Hz to 600 kHz is 55 mrad. To verify the uniformity of the offset frequencies, $\xi$ are measured at two distinct spectral regions other than 1553.5 nm where the beat note is used for stabilization. The selected spectral segments (marked red in Figure 3c) are representative as each $\xi$ is generated from the overlap of different groups of secondary comb lines. Counter results and the corresponding histogram analysis (insets) are summarized in Figures 3d and 3e. The mean values at 1544.72 nm and 1547.86 nm are 523349999.84 Hz and 523349999.92 Hz respectively, while the beat note at 1553.5 nm is stabilized to 523350000 Hz. Offset frequencies at different spectral regions are identical within a sub-Hz error, confirming the uniformity of $\xi$ across the Kerr frequency comb. Phase locking of $f_{rep}$ and $\xi$ to low noise microwave oscillators is complete and it should guarantee the optical frequency stability of Kerr frequency comb.

To quantify the out-of-loop frequency instability of the stabilized Kerr frequency comb, two comb lines (pump at 1598 nm and i$^{th}$ comb at 1555 nm) are compared to the FFC and the heterodyne beat frequencies are counted with a 10-digit, Λ-type frequency counter. The FFC is independently stabilized with the *f-2f* interferometer technique and the frequency fractional instability is $10^{-11}$ at 1 second (Supplementary Information section V). Figure 4a illustrates the more than 20 dB pump frequency noise suppression with both the $f_{rep}$ and $\xi$ phase-locked loops engaged. Figure 4b further shows the Allan deviations (ADs) of the two stabilized Kerr frequency comb lines. A frequency fractional instability of $5.0 \times 10^{-11}/\sqrt{\tau}$, close to the 17.9 GHz reference local oscillator, is measured when the gate time is below 5 second. No apparent difference is observed between the ADs of the two comb lines 43 nm apart, indicating a good coherence transfer across the Kerr frequency comb. For longer gate times, the ADs show a characteristic linear dependence on the gate time that can be attributed to the uncompensated ambient temperature drift. For instance, considering the current chip holder has a long term temperature stability of <10 mK which is limited by the resolution of the temperature sensor, a fractional change of 1.2×10$^{-5}$ in the pump power is necessary to keep the intra-cavity temperature and consequently the $f_{rep}$ constant. Such pump power variation in turn results in a change of 13 kHz in the pump frequency ($\Delta f_p = \frac{\gamma c}{\pi n_g} \Delta P_{int}$ from Eq. 4). The estimated frequency fractional instability is on the order of 7×10$^{-11}$ when referenced to the 188 THz optical carrier, in agreement with the asymptotic behavior of the measured AD.

In summary, we introduce a new Kerr frequency comb stabilization technique which circumvents the requirement of external optical references or power-demanding nonlinear processes. The incommensurate primary and secondary comb line formation of Kerr frequency comb results in an intrinsic offset frequency beat note that can be utilized to gauge the optical frequency fluctuation. Existence of Kerr frequency comb with only one set of primary comb lines, critical for the introduced phase stabilization method, is a universal property of these nonlinear microresonators (Supplementary Information section VI). The sensitivity is measured as $3.7\times10^{-2}$, already more than two orders of magnitude larger than the optical frequency division ratio, and it can be improved by novel microresonator designs to suppress the mode hybridization *(30)*. Furthermore, simultaneous phase locking of $f_{rep}$ and $\xi$ to low noise microwave oscillators guarantees the optical frequency stability of Kerr frequency comb. Out-of-loop AD measurements of the stabilized Kerr frequency comb demonstrates a short term frequency fractional instability of $5\times 10^{-11}/\sqrt{\tau}$, close to the reference microwave oscillator. For gate times longer than 5 second, AD shows a characteristic linear dependence on the gate time that is attributed to the uncompensated ambient temperature drift. Such long term drift can be improved by a better thermal shield or a more effective temperature control *(31, 32)*. Our method preserves the Kerr frequency comb's key advantages of low SWaP and potential for chip-scale electronic and photonic integration, facilitating the already remarkable progress towards chip-scale optical frequency combs for precision spectroscopy, frequency metrology, and optical communication.

**Materials and Methods**

**Si$_3$N$_4$ microresonator fabrication**: First a 3 μm thick oxide layer is deposited via plasma-enhanced chemical vapor deposition (PECVD) on *p*-type 8" silicon wafers to serve as the under-cladding oxide. Then low-pressure chemical vapor deposition (LPCVD) is used to deposit a 725 nm silicon nitride for the spiral resonators, with a gas mixture of SiH$_2$Cl$_2$ and NH$_3$. The resulting silicon nitride layer is patterned by optimized 248 nm deep-ultraviolet lithography and etched down to the buried oxide layer via optimized reactive ion dry etching. The sidewall is observed to have an etch verticality of 85°. Next the silicon nitride spiral resonators are over-cladded with a 3 μm thick oxide layer and annealed at 1200 °C. The refractive index of the silicon nitride film is measured with an ellipsometric spectroscopy from 500 nm to 1700 nm. The fitted Sellmeier equation assuming a single absorption resonance in the ultraviolet, $n(\lambda) = \sqrt{1+\frac{(2.90665\pm 0.00192)\lambda^2}{\lambda^2-(145.05007\pm 1.03964)^2}}$, is imported into the COMSOL Multiphysics for the waveguide dispersion simulation, which includes both the material dispersion and the geometric dispersion.

**Stabilization setup and out-of-loop analysis**: The PI$^2$D control servos we use for feedback in both $f_{rep}$ and $\xi$ phase-locked loops have a full bandwidth of 10 MHz and can be set to have two PI corners, to effectively suppress low frequency noise, in addition to a PD corner to increase the loop stability. To ensure minimal crosstalk between the loops, the PI corners are set at very different frequencies. For the $f_{rep}$ stabilization, the PI corner for the first integrator is set to 200 kHz while the second integrator is switched off. For the $\xi$ stabilization, the PI corners are set to 500 Hz and 50 kHz to achieve higher suppression for low frequency noise. In addition, the PD corners are set to 200 kHz and 100 kHz respectively with a differential gain of 10 dB. The derivative control is important in our system to make the feedback loop more stable and achieve optimal noise suppression. Due to alignment drift in the optics, the mean level of the servo output keeps increasing until the lock is lost in a few minutes. To increase the operation time, we also include in each loop a slow feedback where the feedback error signal is generated by integrating the servo output for 1 second. The control units of the slow feedback loops are the EDFA gain and the PZT, which have larger dynamic ranges than the EOM and the diode current. For out-of-loop analysis, the beat frequency between the Kerr frequency comb and the fiber laser frequency comb is counted with a 10-digit, $\Lambda$-type frequency counter and the Allan deviation is estimated using the equation $\sigma_A(\tau) = \sqrt{\frac{1}{N}\sum_{i=1}^{i=N}\frac{(\bar{y}_{i+1}-\bar{y}_i)^2}{2}}$, where $\tau$, $\bar{y}_i$, and $N = min\left\{20, \left[\frac{300}{\tau}\right]\right\}$ are the gate time, the fractional frequency, and the number of samples respectively. The grating-based filter critically removes the unwanted reference fiber laser frequency comb teeth such that clean heterodyne beat notes with more than 30 dB signal to noise ratio (measured with a 100 kHz RBW), sufficient for reliable counting measurements, can be routinely obtained.

**Supplementary Materials:**
I. Other comb states
II. Device characterization
III. Details of the measurement setup
IV. Derivation of the modulation instability gain peak
V. Out-of-loop characterization
VI. Verification that $\xi$ is a universal property
Figures S1–S7

**Acknowledgements:**

The authors acknowledge discussions with Jinkang Lim, Roberto Diener, Robert Lutwak, and Abirami Sivananthan. **Funding:** This material is based upon work supported by the Air Force Office of Scientific Research under award number FA9550-15-1-0081, and the Defense Advanced Research Projects Agency under award number HR0011-15-2-0014. **Author contributions:** S.W.H. designed the experiment and A.K. conducted the experiment. S.W.H. and J.Y. designed the microresonator. M.Y. and D.L.K. performed the device nanofabrication. S.W.H. and A.K. analyzed the data. S.W.H., A.K., and C.W.W. contributed to writing and revision of the manuscript. **Competing interests:** The authors declare no competing interests. **Data and materials availability:** All data needed to evaluate the conclusions in the paper are present in the paper and/or the Supplementary Materials. Additional data is available from authors upon request.

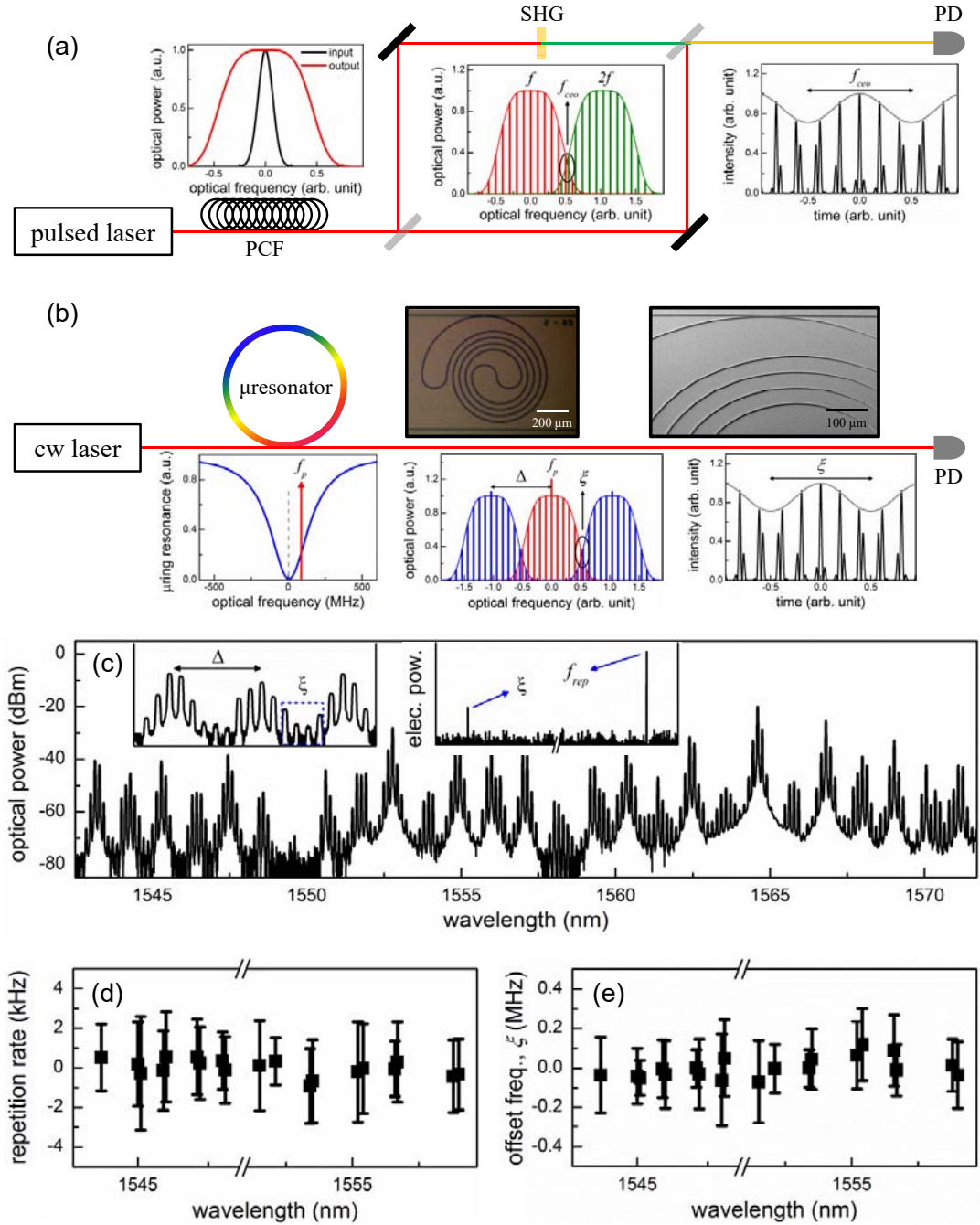

**Figure 1 | Internal comb properties that provide assessment of the optical frequency instability.** (a) Schematic of the current state-of-the-art *f-2f* nonlinear interferometer technique to measure the $f_{ceo}$. By comparing the higher-frequency part of the comb and the second-harmonic generation (SHG) of the lower-frequency end, beat note at $f_{ceo}$ with sufficient signal-to-noise-ratio (SNR) can be generated on a photodetector (PD). The prerequisite of the *f-2f* nonlinear interferometer technique is an octave-spanning comb spectrum, which is typically obtained by spectral broadening of the comb spectrum in a highly nonlinear photonic crystal fiber (PCF). (b) Application of the *f-2f* nonlinear interferometer technique to high-repetition-rate Kerr frequency

comb is challenging due to the requirement of high-peak-power, few-cycle pulses for the nonlinear processes to work properly. On the other hand, the unique generation mechanism of Kerr frequency comb provides an alternative phase stabilization scheme that does not require any external optical reference or power-demanding nonlinear process. Modulation instability and four-wave mixing results in the formation of primary comb lines separated from the pump frequency $f_p$ by $\Delta$ and secondary comb lines spaced by $f_{rep}$. In general, $\Delta$ is not an integer multiple of $f_{rep}$ and thus the Kerr frequency comb exhibits an intrinsic offset frequency, $\xi$. As elaborated later, $\xi$ is linked with $f_p$ and it can also be utilized as the internal comb property to gauge the optical frequency instability. Insets are the optical micrograph and the SEM image of the $Si_3N_4$ microresonator. (c) Kerr frequency comb spectrum in the telecommunication C-band wavelength range. Formation of primary comb lines with $\Delta = 1.1$ nm and overlap between secondary comb lines are observed (left inset). The electrical spectrum of the Kerr frequency comb measures two distinct beat notes of $f_{rep} = 17.9$ GHz and $\xi = 523.35$ MHz (right inset). (d) and (e) To verify that both the comb spacing and offset frequency are uniquely defined across the whole Kerr frequency comb, we measure them at various different spectral segments with a tunable filter (0.22 nm FWHM filter bandwidth). Free-running $f_{rep}$ and $\xi$ at different spectral regions are measured to be the same within error bars of ~2 kHz and ~200 kHz, respectively. At wavelengths where the beat notes have SNR higher than 10 dB (100 kHz RBW), 10 measurements are taken to determine the mean values of the comb spacing and the offset frequencies. The error bar of the measurement is defined as the peak-to-peak deviation from the 10 measurements.

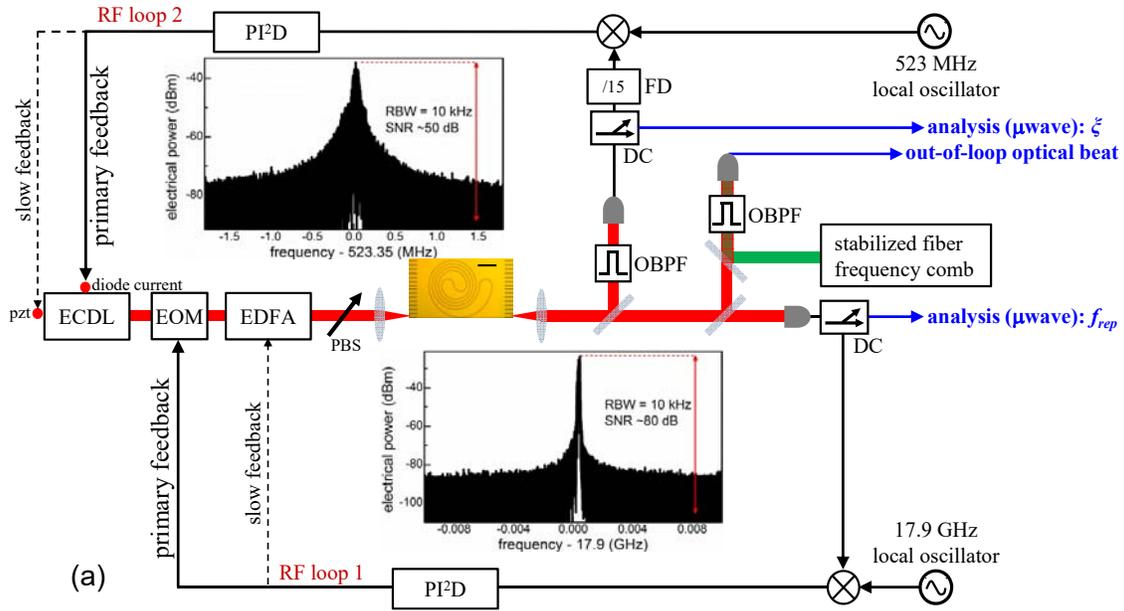

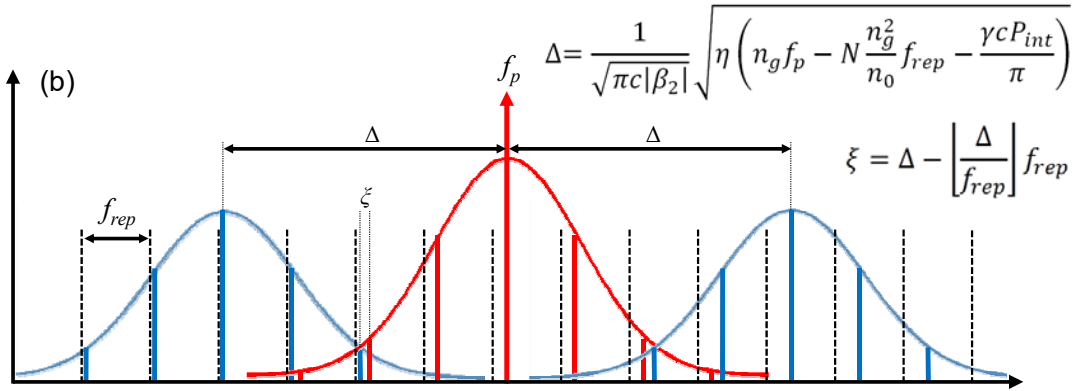

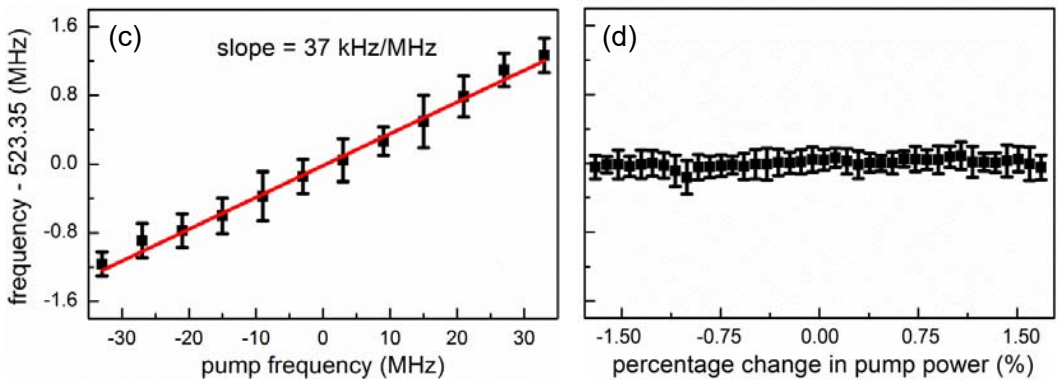

**Figure 2 | Setup of the new phase stabilization technique where the Kerr frequency comb's two degrees of freedom are internally assessed.** (a) Free-running comb spacing and offset frequency are shown in the insets. Phase-locked loops to stabilize the $f_{rep}$ and $\xi$ are engaged with pump power regulation via an electro-optic modulator (EOM) and pump frequency control via diode current of the external cavity diode laser (ECDL), respectively. Additional slow feedbacks

through the gain of the erbium-doped fiber amplifier (EDFA) and the piezoelectric transducer (PZT) of the ECDL are used to extend the stable operation time of the phase-locked loops. Out-of-loop stability of the Kerr frequency comb is evaluated by heterodyne beating with an independent state-of-the-art fiber laser frequency comb (FFC). PBS, polarization beamsplitter; OBPF, optical bandpass filter; DC, directional coupler; FD, frequency divider. (b) Frequency domain illustration of the proposed phase stabilization technique. Here the offset frequency, $\xi$, is linked with the primary comb line spacing, $\Delta$, by the constitutive relation $\xi = \Delta - \left\lfloor \frac{\Delta}{f_{rep}} \right\rfloor f_{rep}$. Furthermore, $\Delta = \frac{1}{\sqrt{\pi c |\beta_2|}} \sqrt{\eta \left( n_g f_p - N \frac{n_g^2}{n_0} f_{rep} - \frac{\gamma c P_{int}}{\pi} \right)}$ where $\beta_2$ is the group velocity dispersion (GVD), $\eta = \frac{\beta_2}{|\beta_2|}$ is the sign of the GVD, $n_g$ is the group index, $n_o$ is the refractive index, $N$ is the longitudinal mode number, $c$ is the speed of light in vacuum, $\gamma$ is the nonlinear coefficient, and $P_{int}$ is the intra-cavity pump power. As $P_{int}$ in the $Si_3N_4$ microresonator is the dominant heat source to change the cavity temperature and subsequently the comb spacing, $f_{rep}$ stabilization in essence eliminates the fluctuation in $P_{int}$. In sum, $\xi$ becomes just a function of $f_p$ after the $f_{rep}$ stabilization. Feedback controls of $f_{rep}$ and $\xi$ thus result in full stabilization of the Kerr frequency comb. (c) Offset frequency as a function of pump frequency, experimentally measured after the $f_{rep}$ stabilization. The pump frequency is stepwise changed by a total of 66 MHz via the PZT of the ECDL. The offset frequency is linearly proportional to the pump frequency with a slope of 37 kHz/MHz. (d) Offset frequency as a function of applied pump power, experimentally measured after the $f_{rep}$ stabilization. The pump power is stepwise changed by a total of 3.4 % via the gain of the EFDA. The offset frequency remains constant within the error bar, verifying the assumption that $f_{rep}$ stabilization effectively eliminates the intra-cavity pump power fluctuation. For (c) and (d), 10 measurements are taken to determine the mean value and the error bars of the measurements are defined as the peak-to-peak deviations from the 10 measurements. Here the pump frequency is not yet stabilized, resulting in the error bars in the offset frequency measurements.

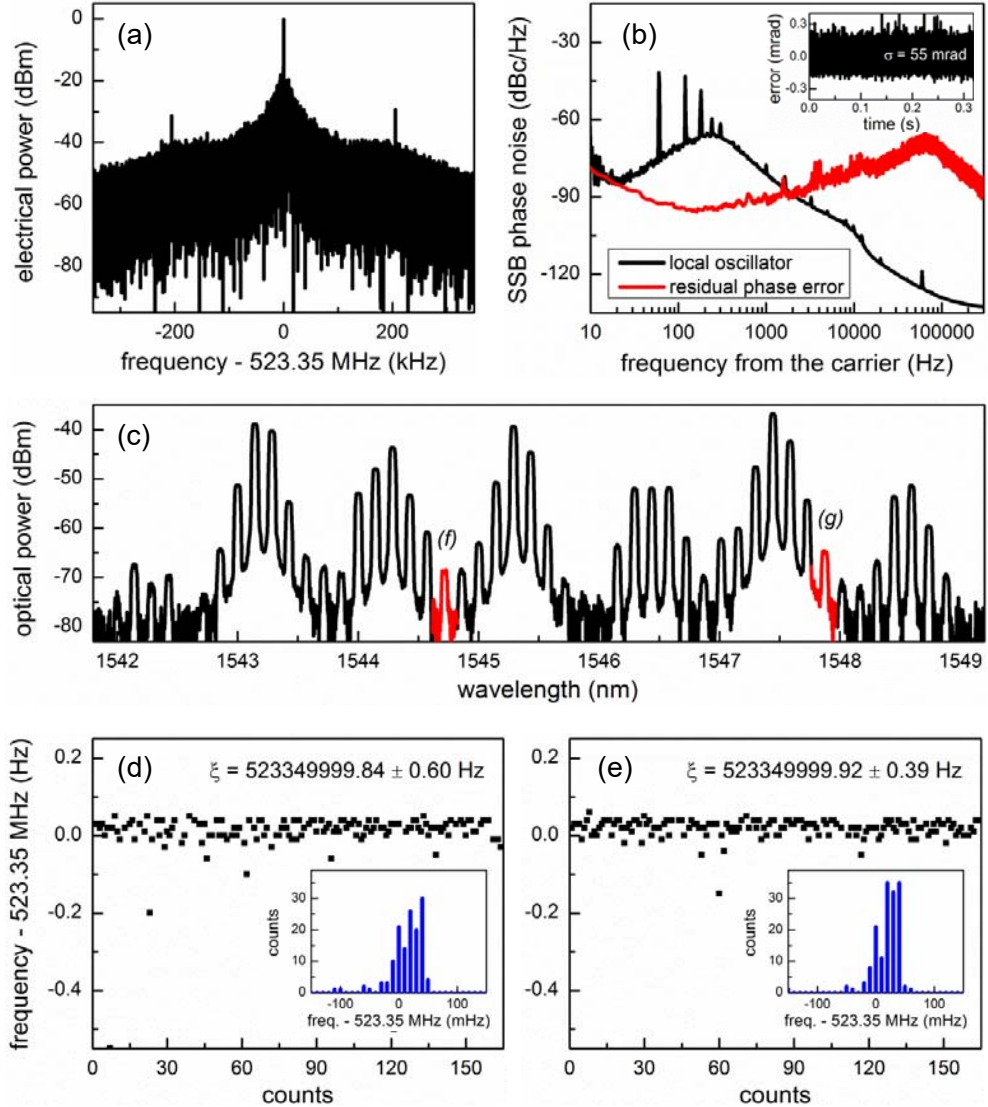

**Figure 3 | Phase stabilization of the Kerr frequency comb without external optical references or power-demanding nonlinear processes.** (a) Electrical spectrum of the stabilized beat note of $\xi$ with an RBW of 10 Hz. To minimize the crosstalk between the two phase-locked loops, here the proportional-integral corner frequency is set lower than that of the $f_{rep}$ loop. Furthermore, a second integrator at 500 Hz and a differentiator at 100 kHz are included to better suppress low frequency noise and improve the loop stability respectively. (b) Single-sideband (SSB) phase noise of the reference 523.35 MHz local oscillator and the residual loop error, showing excess phase noise of the stabilized $\xi$ above 2 kHz from carrier. Inset: rms phase error integrated from 6 Hz to 600 kHz is 55 mrad. (c) To verify the uniformity of the offset frequencies, $\xi$ are measured at two other spectral regions (marked in red; 1544.72 nm and 1547.86 nm) beside the 1553.5 nm region where the beat note is stabilized to 523350000 Hz in the phase-locked loop. The selected spectral segments are representative as each $\xi$ is generated from the overlap of different groups of secondary comb lines. (d), (e) Counter results and the corresponding histogram analysis (insets). The mean value at 1544.72 nm is 523349999.84 Hz,

the standard deviation over 160 measurements is 600 mHz, and the interquartile range is 50 mHz. The mean value at 1547.86 nm is 523349999.92 Hz, the standard deviation over 160 measurements is 390 mHz, and the interquartile range is 40 mHz. Offset frequencies at different spectral regions are identical within a sub-Hz error, confirming the uniformity of $\xi$ across the Kerr frequency comb.

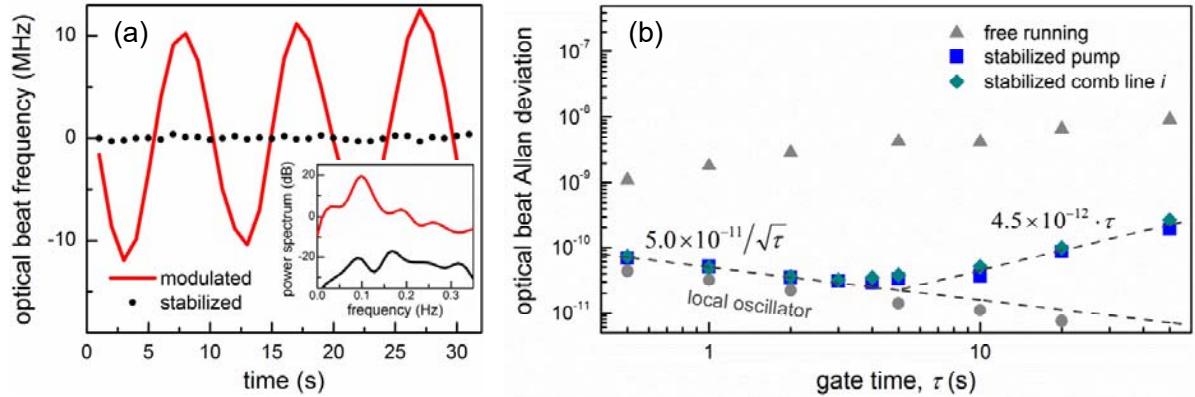

**Figure 4 | Out-of-loop assessment of the internally stabilized Kerr frequency comb.** (a) Optical beat frequency between the pump and the FFC. With both the $f_{rep}$ and $\xi$ phase-locked loops engaged, the artificially introduced pump frequency perturbation (red curve) is indeed greatly suppressed and the optical beat frequency remains constant (black curve). The inset plots the corresponding power spectral densities, showing a more than 20 dB pump frequency noise suppression by the stabilization loops. (b) Allan deviations (ADs) of the stabilized Kerr comb lines, measuring a frequency fractional instability of $5 \times 10^{-11}/\sqrt{\tau}$ when the gate time is below 5 second. For longer gate times, ADs show a characteristic linear dependence on the gate time that can be attributed to the uncompensated ambient temperature drift. No apparent difference is observed between the ADs of the two comb lines 43 nm apart, indicating a good coherence transfer across the Kerr frequency comb.

# Phase stabilization of Kerr frequency comb internally without nonlinear optical interferometry

S.-W. Huang[1*+], A. Kumar[1+], J. Yang[1], M. Yu[2], D.-L. Kwong[2], and C. W. Wong[1*]

[1] Fang Lu Mesoscopic Optics and Quantum Electronics Laboratory, University of California Los Angeles, CA, USA

[2] Institute of Microelectronics, Singapore, Singapore

[*] swhuang@seas.ucla.edu, cheewei.wong@ucla.edu

[+] these authors contributed equally to this work

The Supplementary Information consists of the below sections:

    I.       **Other comb states**

    II.      **Device characterization**

    III.     **Details of the measurement setup**

    IV.     **Derivation of the modulation instability gain peak**

    V.      **Out-of-loop characterization**

    VI.     **Verification that $\xi$ is a universal property**

## Section I: Other comb states

The general multiple mode-spaced (MMS) scheme of comb formation involves the generation of several subcomb families with incommensurate spacing between them *(11)*. This is illustrated in Figure S1a, as we might expect, combs evolving via this scheme would, in general, produce several low frequency RF beats. The comb state we stabilize however, is one with a single offset beat and has just one other subcomb family aside from the sub-comb around the primary comb line as illustrated in Figure S1b. This state is not a necessary part of the comb evolution process and is only observed under the right conditions of power and detuning. Here we briefly describe several other states that we observe in our microresonator. One of the comb states we have observed, generates an equally spaced set of beats spanning around 600 MHz. This 'RF comb' is shown in Figure S2a, in this particular case an interesting point to note is that although multiple subcomb families exist in this state, the RF beats being equally spaced indicates a relationship between the different subcomb families. As detuning is changed this state changes to one with higher noise that doesn't show a regular equally spaced comb structure in the RF domain, as shown in Figure S2b. This state then eventually evolves into one with continuous low frequency noise, the RF spectrum at the repetition rate of such a comb is shown in Figure S2c. In addition, we observe states similar to the one we use for stabilization, having a strong low frequency RF beat in addition to beat due to $f_{rep}$, as shown in Figure S2d, but exhibiting slightly different behavior with regards to degree of correlation between pump and the offset beat.

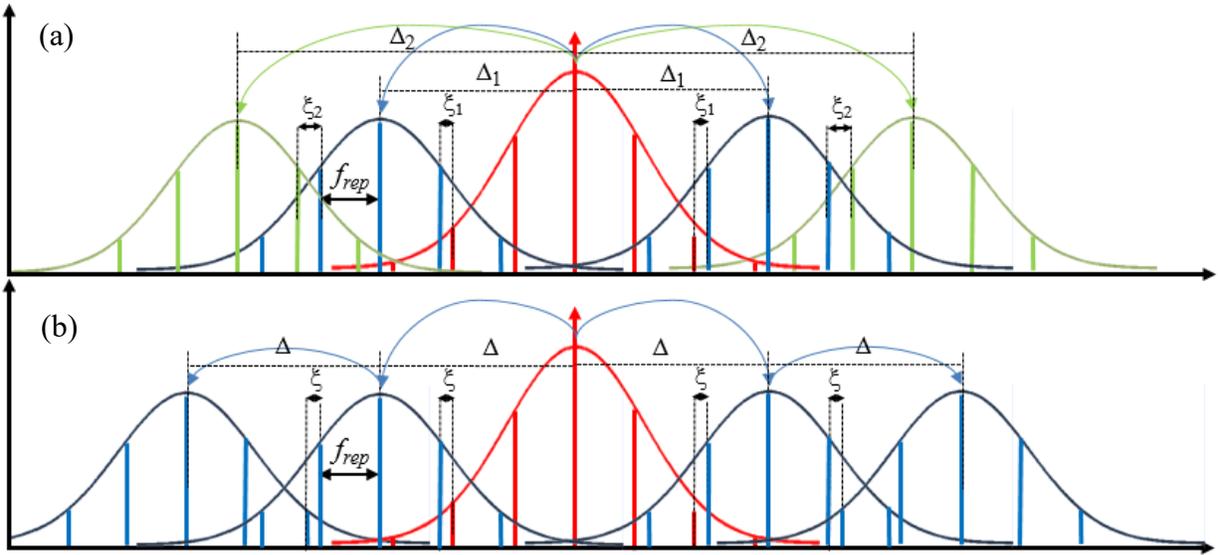

**Figure S1 | a,** The general MMS scheme of comb formation, the two sets of subcombs, shown in blue and green belong to different families because the two sets of primary comb lines around which the subcombs form, are generated independently by the pump. The first set of primary comb lines are formed at an offset of $\Delta_1$ from the pump and the second set are formed at an offset of $\Delta_2$ from the pump, since $\Delta_2$ is not a multiple of $\Delta_1$ and neither $\Delta_2$ nor $\Delta_1$ need be integral multiples of the $f_{rep}$, there are two offset beats generated by beating of subcombs with each other, these offset beats are shown in the schematic as $\xi_1$ and $\xi_2$. Now if this idea is extended to multiple subcomb families we would expect the generation of multiple RF beatnotes, (and if the subcombs were broad enough we would also generate harmonics of the beatnotes) and this is what we experimentally observe. **b,** A special case of MMS comb formation that results in the generation of a single RF beat note (aside from the beat due to $f_{rep}$) that corresponds to the offset $\xi$ between subcombs. Note that, in this case, only the first set of primary comb lines is formed due to modulation instability via the pump, all other primary comb lines are generated via cascaded four-wave mixing between the pump and the first set of primary comb lines, this mechanism allows for a single offset $\xi$, throughout the comb. We choose to stabilize this particular state due to the strong correlation between the pump frequency and $\xi$ due to the dependence of $\xi$ on $\Delta$, as described in the main text.

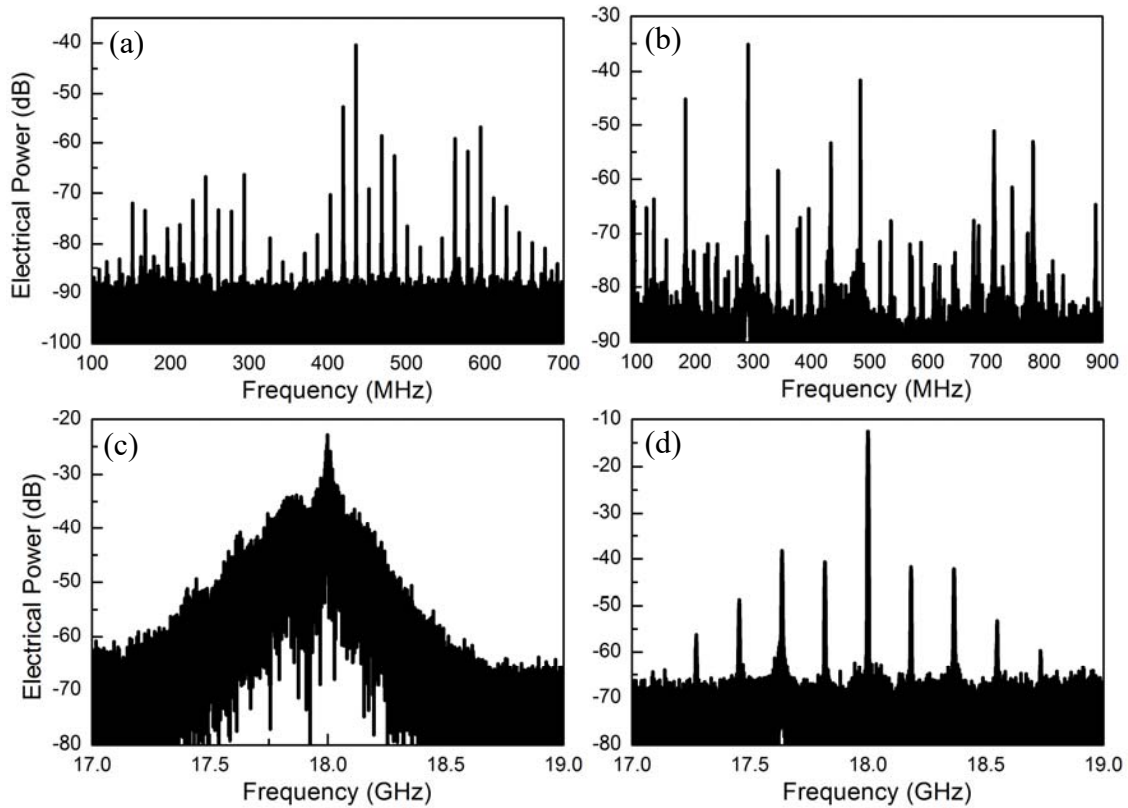

**Figure S2 | a,** Multiple RF beats spanning 600 MHz with a spacing of 16 MHz generated by a comb state. The beats being equally spaced indicates that there is correlation between the offsets of different subcomb families. **b,** Multiple RF beats spanning over a GHz, generated by a comb state. Lack of defined structure to the beats suggests a general MMS scheme for the evolution of the state **c,** RF spectrum at the repetition rate of a comb state showing continuous low frequency noise, this state is obtained from the state in b. by changing the detuning such that the number of RF beats keeps increasing till we eventually have a 'noise pedestal' of continuous noise. **d,** RF spectrum at the repetition rate of a comb state showing a strong offset beat along with multiple harmonics. This state is similar to the one we stabilize; except for the fact that it is less stable to change in pump power or detuning (there is a sudden transition to another state). It also exhibits different behavior with regards to degree of correlation between pump frequency and offset beat.

## Section II: Microresonator dispersion

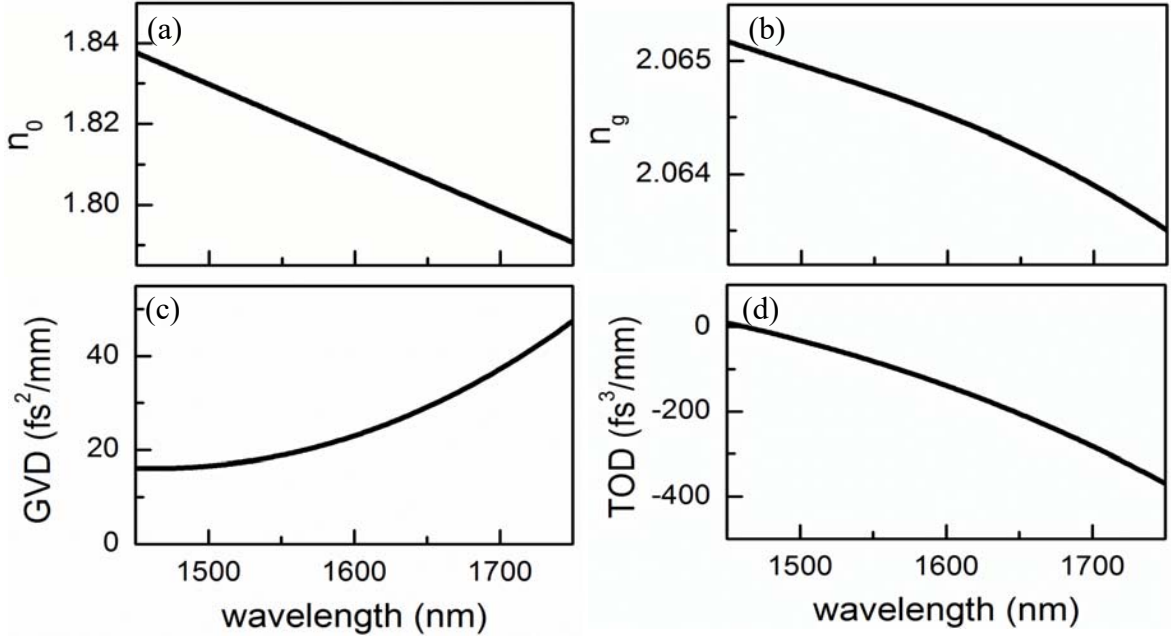

**Figure S3** | Waveguide dispersion is calculated taking into account of both the material dispersion and the geometric dispersion. **a,** Refractive index $n_o$, measured at 1.81 at pump wavelength of 1598 nm. **b,** Group index $n_g$, measured at 2.064 at the pump wavelength. **c,** Group velocity dispersion (GVD) measured at 23 fs$^2$/mm at the pump wavelength. **d,** Third-order dispersion (TOD) measured at 265 fs$^3$/mm at pump wavelength of 1598 nm.

## Section III: Details of the measurement setup

The measurement setup is shown in Figure 2 in the main text. The comb spacing is measured by sending a section of the comb to a high speed photodetector to directly detect the beat note from the repetition rate $f_{rep}$. We then obtain the error signal for feedback by downmixing the output signal with a 17.9 GHz local oscillator. This error signal is the input to a PI$^2$D lock box with a bandwidth of 10 MHz, which sends the feedback signal to an EOM to modulate the input power of the 3W EDFA which pumps the microresonator. The EDFA is operated in the current control mode to achieve effective modulation of the output power. In addition to power modulation via the EOM, we also have a secondary feedback signal (derived by integrating the primary feedback control signal) to the EDFA which directly modulates the power, relatively slowly, primarily with the objective to increase the dynamic range of the lock (EDFA is not used as the sole feedback because it cannot be operated at the full feedback bandwidth). The feedback is designed in the above manner, with fast feedback via the EOM for high feedback bandwidth and slow feedback via the EDFA for high dynamic range, to preserve an optimal lock for a long period of time. Figure S4 summarizes the quality of the $f_{rep}$ stabilization. After the stabilization of the comb spacing, we notice that the offset beat $\xi$ also becomes more stable as can be visually observed from Figure S5. This is per our expectation of partial correlation between $\xi$ and $f_{rep}$ as described in the main text.

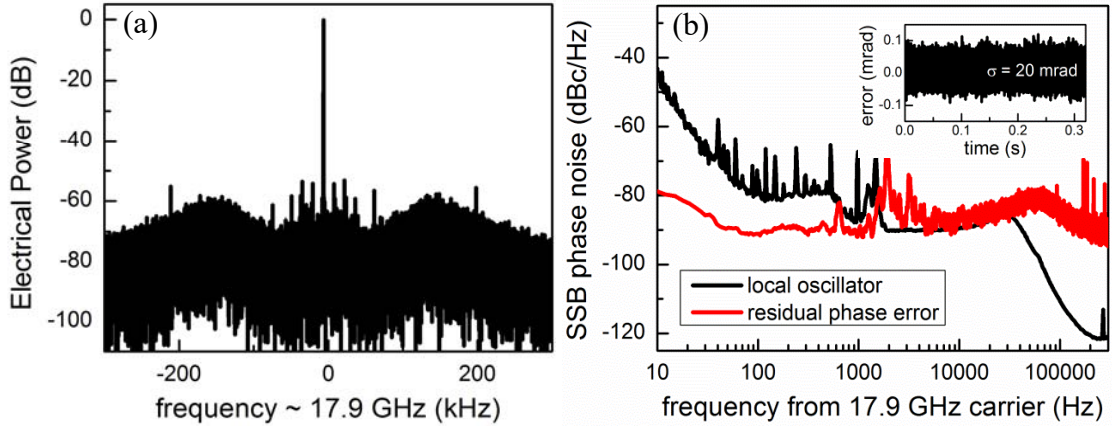

**Figure S4 | a,** RF spectrum of the stabilized beat note of $f_{rep}$ with an RBW of 10 Hz. In the PI$^2$D loop filter, the PI corner and differential frequency were both set at 200 kHz. The design provides a delicate compromise between noise suppression and loop stability. A remaining small noise oscillation at 205 kHz, however, is still present. **b,** Single-sideband phase noise of the reference 17.9 GHz local oscillator and the residual error from the $f_{rep}$ phase-locked loop, showing an excess phase noise of the stabilized comb spacing above 40 kHz from carrier. Inset: rms phase error integrated from 6 Hz to 600 kHz is 20 mrad.

The offset frequency $\xi$ can be used as an indicator of pump frequency after stabilization of $f_{rep}$, as explained in detail in the main text. We therefore use this signal to stabilize the pump frequency when the comb spacing is locked. To achieve a high SNR (which is required to lock $\xi$ effectively), we use an optical grating filter to select a 1-nm section of the comb where the beat frequency is strongest and then send that section to a photodetector to detect the beat (SNR is higher because of a strong beat note in the localized region and also because the detector is not saturated by the $f_{rep}$ beat note, which is much stronger when a larger region of the comb is considered). We send the output to a divide-by-15 frequency divider, and then downmix the signal with a local oscillator operating at 33 MHz to obtain the error signal. The offset beat $\xi$ has more high frequency noise than $f_{rep}$, as we might expect, because it is affected not only by the pump frequency instability but also by high frequency noise in pump power that is not fully compensated by $f_{rep}$ stabilization, the frequency divider is therefore necessary to reduce the high frequency noise and increase the efficacy of the lock. The error signal is sent to a PI$^2$D lock box which provides a feedback signal to modulate the diode current of our ECDL which stabilizes the pump frequency. Similar to the feedback to lock $f_{rep}$, we use a slower secondary feedback (derived from the integrated primary feedback control signal) via the piezo controller of the ECDL to increase dynamic range and preserve the lock for a longer time.

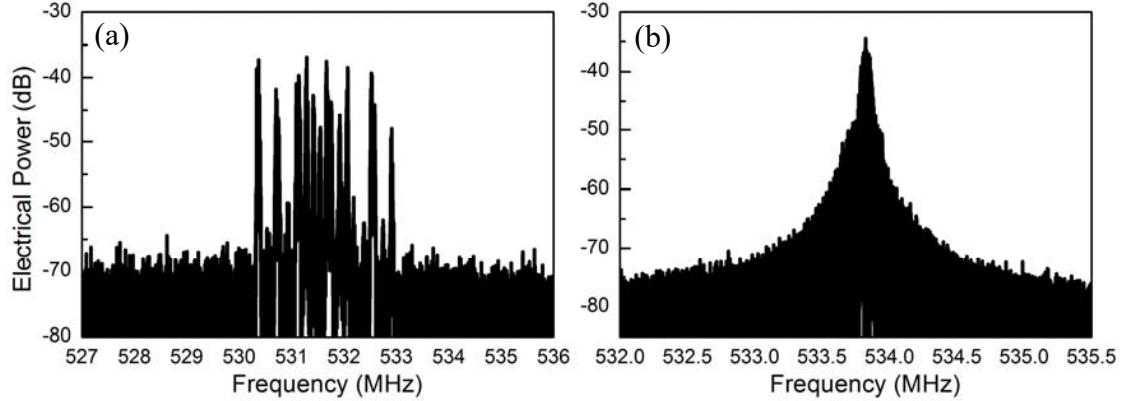

**Figure S5 | a,** The measured offset beat $\xi$ at an RBW of 100 kHz when the $f_{rep}$ is not stabilized. The high noise in the beat arises because the offset frequency $\Delta$ depends on pump frequency and intracavity power given by $\Delta = \frac{1}{\sqrt{\pi c |\beta_2|}} \sqrt{\eta \left( n_g f_p - N \frac{n_g^2}{n_0} f_{rep} - \frac{\gamma c P_{int}}{\pi} \right)}$ and since $\xi = \Delta - \left\lfloor \frac{\Delta}{f_{rep}} \right\rfloor f_{rep}$ fluctuations in both pump frequency and $f_{rep}$ add to instability in $\xi$. **b,** The measured offset beat $\xi$ at an RBW of 10 kHz after $f_{rep}$ is stabilized. We observe an increase in stability of $\xi$ after stabilization of $f_{rep}$ (and thereby stabilization of pump power). Residual noise in the beat note is due to pump frequency noise (and residual noise in pump power after $f_{rep}$ stabilization). $\xi$ can therefore be used to sense pump frequency fluctuations and stabilize it via feedback.

**Section IV: Derivation of the modulation instability (MI) gain peak**

We investigate the intracavity MI gain and derive the frequency at which it is maximum. Eq S1, written below describes the cavity boundary conditions and Eq S2 describes the wave propagation in the cavity when subject to chromatic dispersion and the Kerr nonlinearity *(28)*:

$$E^{n+1}(0,t) = \sqrt{\rho}E^n(L,t)\exp(i\varphi_0) + \sqrt{T}E_i, \qquad (S1)$$

$$\frac{\partial E^n(z,t)}{\partial z} = -i\frac{\beta_2}{2}\frac{\partial^2 E^n(z,t)}{\partial t^2} + i\gamma|E^n(z,t)|^2 E^n(z,t), \qquad (S2)$$

Under the normalization $U^n = \sqrt{\gamma L}E^n$; $\xi = \frac{z}{L}$; $\tau = \frac{t}{\sqrt{|\beta_2|L}}$, the NLSE is reduced to $U^n_\xi = -i\left(\frac{\eta}{2}\right)U^n_{\tau\tau} + i|U^n|^2 U^n$ where $\eta = \frac{\beta_2}{|\beta_2|}$.

Assume steady state continuous wave in the cavity, one such solution is $U^n(\xi,\tau) = U_0\exp(i|U_0|^2\xi)$. A periodic fluctuation generated by instability is modelled by:

$$U^n(\xi,\tau) = [U_0 + v^n_1(\xi) \times \exp(i\Omega\tau) + v^n_{-1}(\xi) \times \exp(-i\Omega\tau)]\exp(i|U_0|^2\xi)$$

Substituting this in the NLSE yields after some algebra that:

$$\frac{\partial v^n_1}{\partial \xi}\exp(i\Omega\tau) + \frac{\partial v^n_{-1}}{\partial \xi}\exp(-i\Omega\tau)$$

$$= \frac{i\eta\Omega^2}{2}\left(v^n_1\exp(i\Omega\tau) + v^n_{-1}\exp(-i\Omega\tau)\right)$$

$$+ i\left(\left(|U_0|^2 v^n_1 + U_0^2 v^{n*}_{-1}\right)\exp(i\Omega\tau) + \left(|U_0|^2 v^n_{-1} + U_0^2 v^{n*}_1\right)\exp(-i\Omega\tau)\right)$$

This then can be written as:

$$\frac{\partial}{\partial \xi}\begin{pmatrix} v^n_1 \\ v^{n*}_{-1} \end{pmatrix} = \begin{bmatrix} \frac{i\eta\Omega^2}{2} + i|U_0|^2 & iU_0^2 \\ -i(U_0^*)^2 & \frac{-i\eta\Omega^2}{2} - i|U_0|^2 \end{bmatrix}\begin{pmatrix} v^n_1 \\ v^{n*}_{-1} \end{pmatrix}$$

The general solution to the equation above can be written as:

$$\begin{pmatrix} v^n_1 \\ v^{n*}_{-1} \end{pmatrix} = \begin{pmatrix} a^n \\ b^n \end{pmatrix}\exp(\mu\xi) + \begin{pmatrix} c^n \\ d^n \end{pmatrix}\exp(-\mu\xi)$$

With eigenvalue $\mu = \Omega\sqrt{-\eta|U_0|^2 - \Omega^2/4}$ and eigenvector components satisfying:

$$\frac{a^n}{b^n} = \frac{-(U_0)^2}{\left(\frac{\eta\Omega^2}{2} + |U_0|^2 + i\mu\right)}; \quad \frac{d^n}{c^n} = \frac{-(U_0^*)^2}{\left(\frac{\eta\Omega^2}{2} + |U_0|^2 + i\mu\right)}$$

Now including the cavity boundary conditions (by substituting $E_n$ in Eq S1) and noting that $E_i$ is a constant, we see that we can write:

$$v_{\pm 1}^{n+1}(\xi = 0) = \sqrt{\rho} exp(i\varphi_0 + i|U_0|^2) v_{\pm 1}^n(\xi = 1)$$

Since

$$\begin{pmatrix} v_1^n \\ v_{-1}^{n*} \end{pmatrix} = \begin{bmatrix} \left(\frac{\eta\Omega^2}{2} + |U_0|^2 + i\mu\right) & -(U_0^*)^2 \\ -(U_0)^2 & \left(\frac{\eta\Omega^2}{2} + |U_0|^2 + i\mu\right) \end{bmatrix} \begin{pmatrix} a^n \\ c^n \end{pmatrix}$$

And

$$\begin{pmatrix} v_1^{n+1} \\ v_{-1}^{n+1*} \end{pmatrix} = \sqrt{\rho} \begin{bmatrix} exp(i\varphi_0 + i|U_0|^2) & exp(i\varphi_0 + i|U_0|^2) \\ exp(-i\varphi_0 - i|U_0|^2) & exp(-i\varphi_0 - i|U_0|^2) \end{bmatrix} \begin{pmatrix} v_1^n \\ v_{-1}^{n*} \end{pmatrix}$$

We can write after substituting in:

$$\begin{pmatrix} a^{n+1} \\ c^{n+1} \end{pmatrix} = \frac{\sqrt{\rho}}{|s|^2 - t^2} \begin{bmatrix} |s|^2 e^\mu e^{i\vartheta} - t^2 e^\mu e^{-i\vartheta} & |s|^2(e^{-\mu} e^{i\vartheta} - e^{-\mu} e^{-i\vartheta}) \\ t^2(e^\mu e^{i\vartheta} - e^\mu e^{-i\vartheta}) & |s|^2 e^\mu e^{i\vartheta} - t^2 e^\mu e^{-i\vartheta} \end{bmatrix} \begin{pmatrix} a^n \\ c^n \end{pmatrix}$$

, where $s = -(U_0)^2$; $t = \frac{\eta\Omega^2}{2} + |U_0|^2 + i\mu$; $\vartheta = \varphi_0 + |U_0|^2$.

Taking the determinant of this matrix we arrive at the eigenvalues as:

$$q_\pm = \sqrt{\rho}\left(p \pm \sqrt{p^2 - 1}\right) \quad \text{where } p = \cos(\vartheta)\cosh(\mu) - \frac{(2|U_0|^4 + 2i\mu|U_0|^2 + i\mu\eta\Omega^2)}{(2\mu|U_0|^2 + \mu\eta\Omega^2)}\sinh(\mu)\sin(\vartheta)$$

Under the mean field approximation, which in this case implies that $\Omega, |U_0|^2 \sim O(\varepsilon)$ we can approximate this expression to:

$$p = \cos(\vartheta)\cosh(\mu) - \left(|U_0|^2 - \frac{\eta\Omega^2}{2}\right)\frac{\sinh(\mu)}{\mu}\sin(\vartheta)$$

In the good cavity limit that $\delta, T \sim O(\varepsilon)$, we then see that $\Omega_{opt} = \sqrt{2\eta(\delta - 2|U_0|^2)}$.

Now $\delta$ can be written as $\frac{2\pi(f_p - f_o)}{f_{rep}}$ where $f_p$ is the pump frequency, $f_{rep}$ is the comb spacing and $f_o$ is the resonance frequency, $f_{rep}$ can also be expressed as $c/(n_g L)$ and $f_o = N\frac{n_g}{n_o}f_{rep}$ where $n_g$ is the group index and $n_o$ is the refractive index. Furthermore $\omega = \Omega/\sqrt{|\beta_2|L}$ where $\omega$ is the frequency with respect to real time coordinates and $|U_0|^2 = \gamma L|E_o|^2$ where $|E_o|^2$ is the intracavity power, denoted by $P_{int}$.

Putting these together we have:

$$\omega_{opt} = \sqrt{\frac{4\pi n_g \left(f_p - N\frac{n_g}{n_0}f_{rep}\right)}{\eta|\beta_2|c} - \frac{4\gamma P_{int}}{\eta|\beta_2|}}$$

**Section V: Out-of-loop characterization**

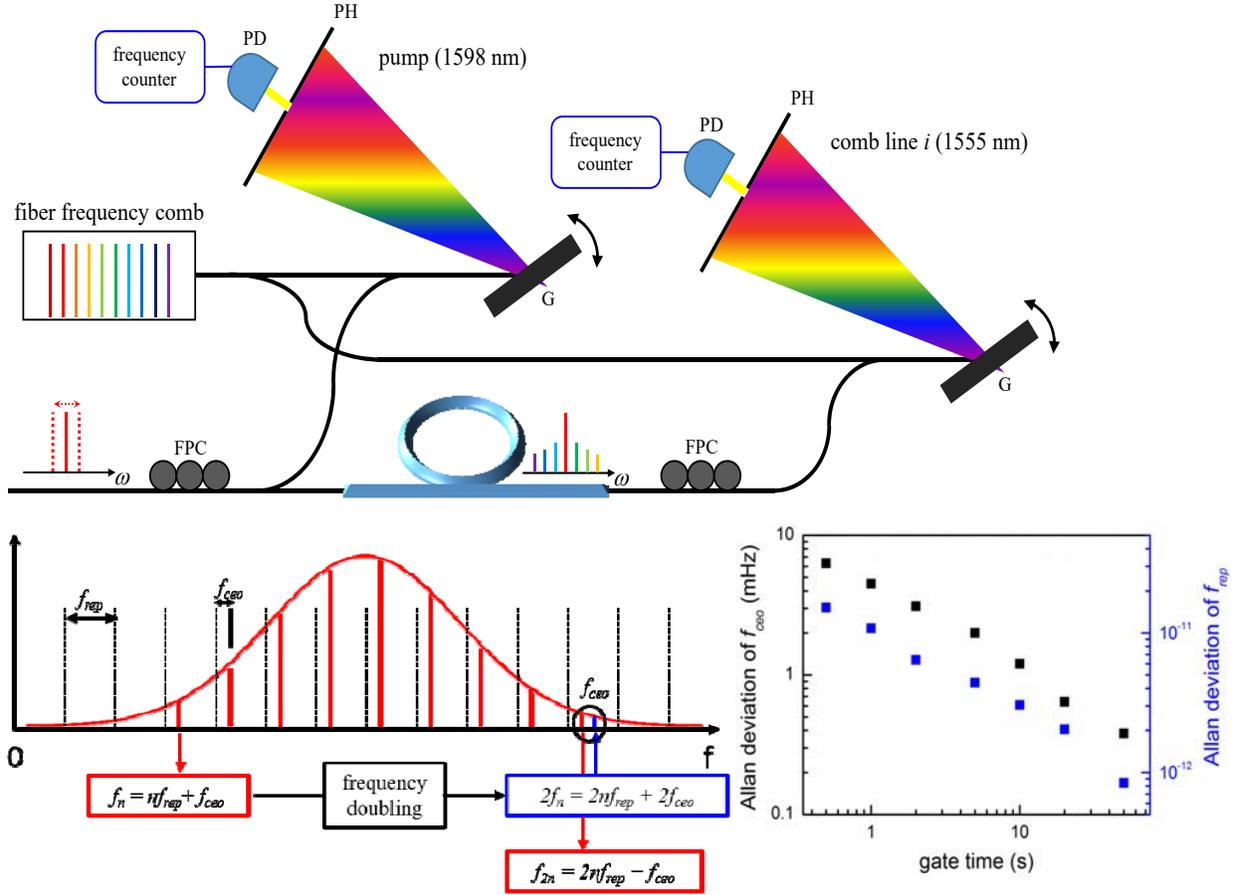

**Figure S6 | a,** To quantify the frequency instability of the Kerr frequency comb, two comb lines (pump at 1598 nm and $i^{th}$ comb at 1555 nm) are compared to an independently stabilized FFC and the heterodyne beat frequencies are counted with a 10-digit, Λ-type frequency counter. The FFC is referenced to a rubidium-disciplined crystal oscillator with a frequency fractional instability of $5\times10^{-12}$ at 1 second. The gratings critically remove the unwanted reference FFC comb lines for reliable counting measurements. **b,** The repetition rate of the FFC (~250 MHz) is detected with a PD and locked to an RF local oscillator, in addition, *f-2f* interferometry is used to detect $f_{ceo}$ and lock it to an RF reference with the same clock as that used to lock $f_{rep}$. **c,** The Allan Deviation of the $f_{ceo}$ is plotted for the FFC in mHz and the Allan Deviation for $f_{rep}$ is plotted relative to the carrier. As we observe from the plot, $f_{rep}$ is the limiting factor for the stability of our reference, which is per our expectation, because of the high sensitivity of the comb line frequencies to $f_{rep}$ due to the low optical division ratio (~$10^{-6}$).

## Section VI: Verification that $\xi$ is a universal property

Formation of Kerr combs with a single offset beat in addition to $f_{rep}$ is not a unique property dependent on microresonator characteristics but is universal and arises from the mechanics of Kerr comb generation. In addition to the case shown in the main text, we also observe a similar state in a single mode $Si_3N_4$ microresonator cavity with a tapered structure *(30)* thereby verifying the universality of such comb modes. The comb spectrum is shown in Figure S7a. and the offset beat is measured in different comb slices, as shown in Figure S7b, to verify that it is truly a single offset comb state.

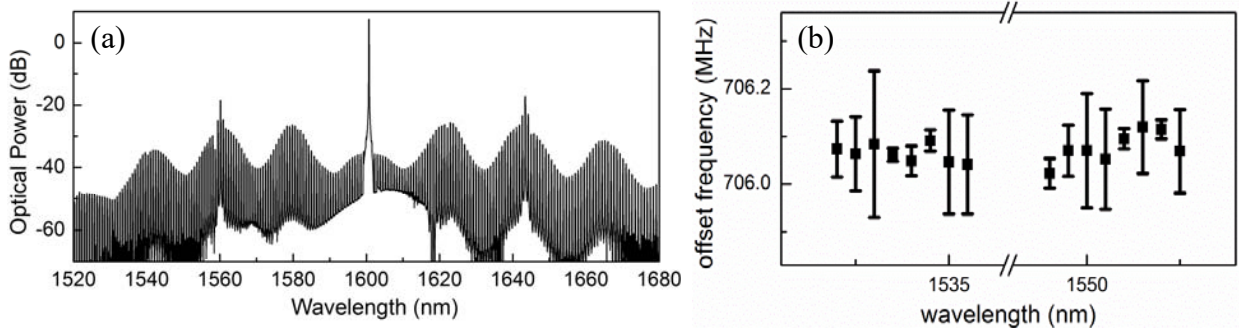

**Figure S7 | a,** Spectrum of a comb-state, similar to the one we stabilize, generated in a single mode microresonator with a tapered structure. This comb state generates a single offset beat $\xi$ in the RF domain in addition to the repetition rate. **b,** To verify that the offset frequency is uniquely defined across the whole Kerr frequency comb, we measure it at various different spectral segments with a tunable filter (0.22 nm FWHM filter bandwidth). Free-running $\xi$ without $f_{rep}$ stabilization (~ 700 MHz) in different spectral regions is measured to be the same within error bars of ~200 kHz. At wavelengths where the beat notes have SNR higher than 10 dB (100 kHz RBW), 10 measurements are taken to determine the mean value of the offset frequency. The error bar of the measurement is defined as the peak-to-peak deviation from the 10 measurements.